\documentclass[aps,prb,preprint,showpacs,superscriptaddress]{revtex4}
\usepackage{graphicx}
\usepackage{hyperref}
\usepackage{natbib}
\begin{document}
\title{Counterintuitive transitions in the multistate Landau-Zener problem with linear level crossings.}
\author{N.A. Sinitsyn$^{}$}
\affiliation{$^{}$Department of Physics, Texas A\&M University, College Station, \\
Texas 77843-4242} 
\date{\today}


\begin{abstract}
We generalize the Brundobler-Elser hypothesis in the multistate Landau-Zener problem to the case when instead of a state with
the highest slope of the diabatic energy level there is a band of
states with an arbitrary number of parallel levels having the same slope. We argue that the probabilities of counterintuitive
transitions among such states are exactly zero. 
\end{abstract}
\maketitle

\pagenumbering{arabic}

\begin{figure}
\includegraphics[width=8cm]{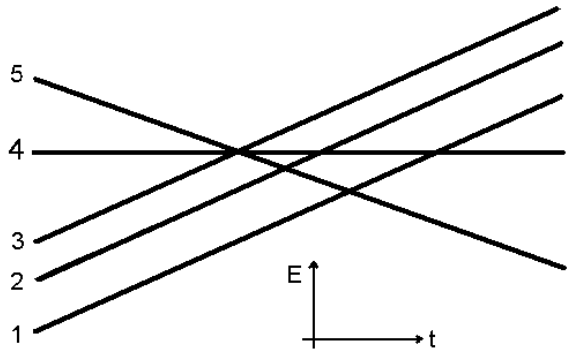}
\caption{Diabatic energies of a 5-state Landau-Zener model. The choice of parameters is as follows, $\beta_1=\beta_2=\beta_3=1$, $\beta_4=0$, 
$\beta_5=-0.8$, $\alpha_1=0$, $\alpha_2=0.3$, $\alpha_3=0.5$, $\alpha_4=0$, $\alpha_5=0.4$.}
\label{LZfig1}
\end{figure}

  The multistate Landau-Zener problem has been an active field of research during the last decade with
  various applications in condensed matter and atomic physics. The two state problem with linear time-dependence of diagonal elements of the
  Hamiltonian was solved exactly by Landau and Zener \cite{{Landau},{Zener}}. The simplest 
  generalization of the two state problem is the Schr\"odinger equation of the form
  \begin{equation}
   i\dot{\psi}(t)=(A+Bt)\psi (t)
   \label{mlz}
   \end{equation}
  where
  $A$ and $B$ are Hermitian $N \times N$ matrices with constant elements. The Hamiltonian of the model is $H=A+Bt$. 
  The matrix $B$ can be always chosen diagonal. The goal of the theory is to find 
  the transition probabilities, namely the squared elements of the scattering matrix $\lim \limits_{t^{\prime} 
  \rightarrow +\infty\, ,t   \rightarrow -\infty} |S_{ij}(t^{\prime},t)|^2$ where $i$ and $j$ enumerate eigenstates (the so called diabatic states)
  of the matrix $B$ with time-dependent diabatic energies $E_i(t) = B_{ii}t+A_{ii} \equiv \beta_i t + \alpha_i$. Nondiagonal elements
  of the matrix $A$ that couple diabatic states with the same slopes $\beta_i$ can always  be made zero by a change of the basis. 
  It is convinient to visualize the time-dependence of diagonal elements of any such model in the time-energy diagram like the one in Fig.\ref{LZfig1}.
  When all crossing points are well separated one can try to solve the problem naively by a successive application of the two state Landau-Zener formula
  at every two level intersection. Even in this approximation the dependence of transition probabilities on parameters can be very complicated since amplitudes
  of different paths
  leading to the same final states can interfere. The task becomes even more complicated when more than two levels can be close to each other 
  simultaneously. Then even approximate estimates become very sophisticated \cite{Gefen}. In spite of this complexity, there have been
  a number of remarkable efforts to solve the model (\ref{mlz}) exactly at least for some special choices of parameters. Generally this requires 
  nontrivial approaches because to solve the n-state model one must consider a n-th order differential equation with time-dependent
  coefficients. 
  
  Although a few important
  classes of exactly solvable models of
  the type (\ref{mlz}) have been known for a long time \cite{{demkov},{zeeman2}} the interest toward exact results in the multistate Landau-Zener problem has grown
  up after the work of 
  Brundobler and Elser \cite{brund}, who noticed that for any model of the form   (\ref{mlz}) there are elements of the transition probability matrix
  that can be found by a simple application of the two state Landau-Zener formula at every intersection of diabatic energies. Particularly, they presented
  an empirical formula for the diagonal element of the scattering matrix for the state whose diabatic energy level has the highest slope, i.e. if
  $k$ is the index of the state with $\beta_k=\max(\beta_1 \ldots \beta_N)$ or $\beta_k=\min(\beta_1 \ldots \beta_N)$ then
  \begin{equation}
  |S_{kk}(+\infty,-\infty)|=exp \left( - \pi \sum \limits_{i\,(i \ne k)} \frac{|A_{ki}|^2}{|\beta_k-\beta_i|} \right)
  \label{eq2}
  \end{equation}
  The formula (\ref{eq2}) is confirmed by all known exactly solvable models 
  with finite number of states \cite{{demkov},{zeeman2},{bow},{dem33},{usuki},{sinitsyn}} and by multiple numerical checks. 
  The authors of \cite{brund}
  speculated that this finding probably indicates that the whole problem (\ref{mlz}) can be solved exactly or at least can be understood in terms of 
  the two-level crossings. Various exact solutions and approximations seem support this idea \cite{{sinitsyn},{math}}.  Recent work \cite{shytov}
  demonstrated that (\ref{eq2}) follows from a simple analytical continuation of the asymptotic solution into the complex time, though such a 
  procedure fails to predict correctly other elements of the scattering matrix.
  The goal of the present work is to demonstrate that the Brundobler-Elser hypothesis can be generalized to some nondiagonal elements of
  the scattering matrix and to explain why analytical continuation of amplitudes into the complex times provides correct predictions for 
  some elements of the scattering matrix.
  
  Assume that instead of one state with the highest slope of diabatic energy level there is a band of an arbitrary number of states having the same 
  highest slope  so that diabatic energies in this band are different only by constant parameters $\alpha_m$.
  If we assume a "semiclassical" approximation where transition between any two states happen only at the corresponding crossing point of 
  their diabatic energies then there are 
  elements of the transition probability matrix that would be zero in this approximation. Such transitions, if happen, are called 
  counterintuitive transitions \cite{Yur}. Thus, in the model shown in Fig.1, transitions from the state $1$ to states $2$ and $3$ and from the state $2$ to the state
  $3$ are counterintuitive.

   Generally for the model (\ref{mlz}), if $\beta_m=\beta_n=\max(\beta_1 \ldots \beta_N)$
   then the transition from the state  $m$ to the state 
  of the same band $n$ would be counterintuitive if  $\alpha_m<\alpha_n$. Correspondingly, if 
  $\beta_m=\beta_n=\min(\beta_1 \ldots \beta_N)$ then the transition is counterintuitive if  $\alpha_m>\alpha_n$. 
  We argue that in the multistate Landau-Zener model with linear time-dependence of diabatic energies such
  counterintuitive transitions have exactly zero probability i.e. without assuming any semiclassical approximation for any model of the type 
  (\ref{mlz}), if the transition from the state $m$ to the state $n$ is counterintuitive, then 
  \begin{equation}
  |S_{nm}(+\infty,-\infty)|=0
  \label{eq3}
  \end{equation}

  
  The "no-go" formula (\ref{eq3}) and the Brundobler-Elser conjecture (\ref{eq2}) can be 
  understood by the approach similar to the one used by Landau in the two state calculations \cite{Landau}. Since we are interested in the asymptotic
  magnitude of the amplitudes we can analytically extend the evolution (\ref{mlz}) to imaginary time and choose the evolution path so that 
  always $|t|\rightarrow \infty$. The distances between instantaneous eigenenergies $\epsilon _i(t)$ of the Hamiltonian remain always large in this case,
  namely of the order
  of $|(\beta_i-\beta_j)t|>>|A_{ij}|$ for the states $i \ne j$ and hence we can use the adiabatic approximation
  \begin{equation}
  \psi_i(t)=e^{-i\int_{t_0}^{t} \epsilon _i(t)dt}\psi_i(t_0)
  \label{adiab}
  \end{equation}
  where  the state $\psi_i$ has the leading asymptotic $\psi_i \sim \exp(-i\beta_it^2/2)$ at $t\rightarrow -\infty$.
  
\begin{figure}
\includegraphics[width=8cm]{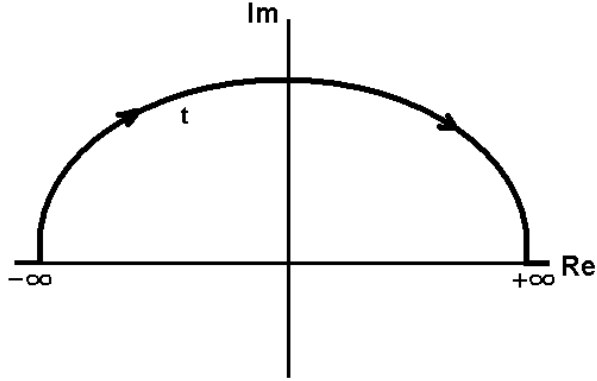}
\caption{The deformed time contour for the evolution from large negative to large positive times with $t=R \exp(i\phi)$, $R\rightarrow \infty $,
$0  \le \phi \le \pi$. }
\label{LZfig2}
\end{figure}

  The approximation (\ref{adiab}) becomes exact in the limit $t \rightarrow \infty$ but it
  is valid generally only if there are no other solutions that become exponentially large in comparison with the state $\psi_i$ to which it is applied. 
  Suppose that the state 
  $\psi_0$ has the largest slope of the diabatic energy $\beta_0$ at $t\rightarrow - \infty$ and is initially occupied. 
  In this case it is convenient to choose the time-path as shown in Fig.\ref{LZfig2} with $t=R \exp(i\phi)$ where $R \rightarrow \infty$ and 
  $\phi$ decreases from $\pi$ to zero.
  One can always change variables so that $\beta_0=0$ 
  and $\beta_i <0$
  for states with slopes $\beta_i \ne \beta_0$ \cite{brund}. When $\phi$ changes in the interval from 
  $3\pi/4$ to  $\pi/4$, the amplitudes of states with slopes $\beta_i<0$ are decreasing exponentially and become suppressed by the factor
   $\exp( C(\phi) \beta_i |t|^2/2)$ where $\beta_i<0$ and $C(\phi)$ is a positive coefficient that depends only on the angle. 
   We choose the asymptotics so that 
   at the angle $\phi=3 \pi /4$ the state $\psi_0$ is dominating over all others, i.e. is exponentially large in comparison to them.
   Then the states with $\beta_i<0$
   should not affect the adiabatic approximation in the interval $3\pi/4 >\phi > \pi/4$ since they can only decrease there.
   One can see that the condition that at $\phi=3\pi/4$ the state $\psi_0$ is dominating also leads to the vanishing of the 
   amplitudes of other states with $\beta_i<0$ in the interval  $\pi<\phi<3\pi/4$ so that it is not forbidden to choose $|\psi_0(-\infty)|=1$
   and $|\psi_i (-\infty)| \rightarrow 0, \,(i\ne 0)$. 
   
   At the last part of the contour $\pi/4>\phi>0$ amplitudes of states with $\beta_i<0$ grow from almost zero value, 
   but at $\phi=0$ time becomes real and hence amplitudes cannot be larger than unity. So in this part of the contour such amplitudes have not
   enough time to become exponentially large. It means that they still remain small or comparable with $\psi_0$ at this interval
   and the formula (\ref{adiab}) should be 
   valid for the state $\psi_0$ during the whole evolution. Substituting the energy up to the first order correction in $1/|t|$ 
   \begin{equation}
   \epsilon_0 (t) \sim \alpha_0 + \sum \limits_i \frac{|A_{i0}|^2}{(\beta_0 - \beta_i)t}
   \label{en}
   \end{equation}
   into the formula for the transition probability
   \begin{equation}
   |S_{00}|^2=\frac{|\psi_0(+\infty)|^2}{|\psi_0(-\infty)|^2}=exp \left( -2Im(\int \limits _C \epsilon_0 (t) dt)  \right)
   \label{prob}
   \end{equation}
   we find the Brundobler-Elser result (\ref{eq2}).
    It is clear from this analysis why the formula (\ref{eq2}) is generally not valid for other diagonal elements of the scattering matrix. If 
    an initially filled state does not have the highest slope of the energy level there are states with higher slopes that grow exponentially and
    become large in the interval $3\pi/4>\phi>\pi/4$ of the contour so that the adiabatic approximation becomes invalid in
    application to $\psi_0$. To treat this case properly, one 
    should investigate the Stokes phenomenon near all crossing points of diabatic energies \cite{math}.

  This analysis becomes more complicated if there is more than one state having the same largest energy  slope $\beta_0$. 
  If such states have also larger constant part of the
  diabatic energy $\alpha_m>\alpha_0$ they can grow in the first half of the contour as $\exp (C^{\prime}(\phi)\alpha_m |t|)$, i.e faster then the
  initially filled state $\psi_0$,
  but being initially vanishing, amplitudes of such states can grow only due to transitions from the other states.
  At first half of the time-contour they are coupled only to states that are suppressed by much stronger exponents $\exp(C(\phi)\beta_i|t|^2/2)$,
  $(\beta_i<0)$; therefore we do not expect
  that they become large in comparison with $\psi_0$ up to $\phi = \pi/2$. In the second part of the path $\pi/2<\phi<0$ states with
  such an asymptotic  $\exp (-i\alpha t)$ already decrease exponentially and become suppressed in comparison with $\psi_0$;
  therefore we can expect that they do not break the approximation (\ref{adiab}) for the state $\psi_0$ 
  and have vanishing amplitudes at the end of the evolution. 
  This is exactly in agreement with (\ref{eq3}). 
  
\begin{figure}
\includegraphics[width=8cm]{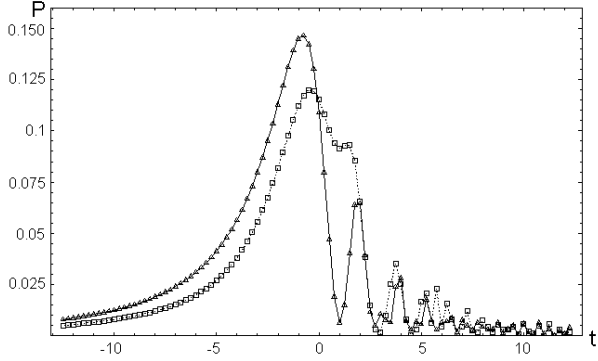}
\caption{Time dependence of the counterintuitive transition probabilities for the model in Fig.\ref{LZfig1}. Triangles
correspond to $P(t)=|S_{21}(t,-\infty)|^2$ and boxes show $P(t)=|S_{31}(t,-\infty)|^2$. The choice of nondiagonal elements of the Hamilton operator
is $H_{12}=H_{13}=H_{23}=0$, $H_{34}=0.8$, $H_{35}=0.3+0.24i$, $H_{24}=0.1+0.7i$, $H_{25}=0.5+0.1i$, $H_{14}=0.4+0.12i$, $H_{15}=0.25+0.2i$, $H_{45}=
0.6+0.9i$. The other elements are obtained by employing Hermitian properties of the matrix $H$.}
\label{LZfig3}
\end{figure}

  Our arguments in support of (\ref{eq2}) and (\ref{eq3}) are certainly very intuitive and every step in the mathematically rigorous proof 
  requires more detailed justification. However, we note that 
   (\ref{eq3}) is also confirmed by
  all known exactly solvable classes featuring the possibility of counterintuitive transitions, namely by the Demkov-Osherov model \cite{demkov},
  the generalized bow-tie model \cite{dem33} and the model of two crossing bands of parallel levels \cite{usuki}. Besides, we performed a number of
  numerical simulations with arbitrary choices of parameters. As we found, all they support our hypothesis (\ref{eq3}). For example, 
  in Fig.\ref{LZfig3} we show the
  time-dependence of the probabilities to find the system at states $2$ and $3$ in the model demonstrated scematically in Fig.\ref{LZfig1}
  if initially
  only the state
  $1$ is occupied. One can deduce that generally during the evolution these probabilities can be rather high ($>0.1$) and show oscillating behavior,
  but asymptotically
  at $t \rightarrow +\infty$ they vanish. Numerically we can simulate the evolution only in the finite time interval. For the evolution from $t=-500$
  to $t=500$ and the same parameters as in Fig.\ref{LZfig1} we find $|S_{21}|^2=5.18 \times 10^{-7}$ and $|S_{31}|^2=3.11 \times 10^{-7}$. 
  In comparison $|S_{11}|^2=0.234$, $|S_{41}|^2=0.295$ and $|S_{51}|^2=0.472$.
  We also note that although counterintuitive transitions
  have vanishing probabilities, the presence of the states $2$ and $3$ does affect other elements of the scattering matrix. Thus if we 
  set all couplings of states $2$ and $3$ with all other states to zero, then numerically 
  calculated nondiagonal transition probabilities are $|S_{41}|^2=0.672$ and $|S_{51}|^2=0.094$ that
  is different from our previous numerical result. 

\begin{figure}
\includegraphics[width=8cm]{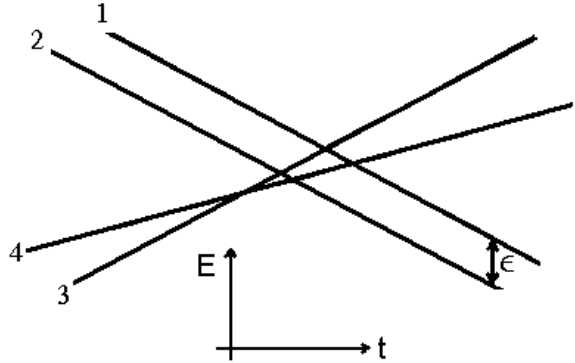}
\caption{Diabatic levels of a 4-state model. The matrix elements of the Hamilton operator are chosen as follows.
$H_{11}=-t$, $H_{22}=-t-\epsilon$, $H_{33}=t$, $H_{44}=0.5t-0.5$, $H_{12}=0$, $H_{13}=0.4-0.1i$, $H_{14}=0.6$, $H_{23}=0.4+0.5i$, $H_{24}=0.2+0.3i$.
The other elements are obtained by employing Hermitian properties of the matrix $H$.}
\label{LZfig4}
\end{figure}

\begin{figure}
\includegraphics[width=10cm]{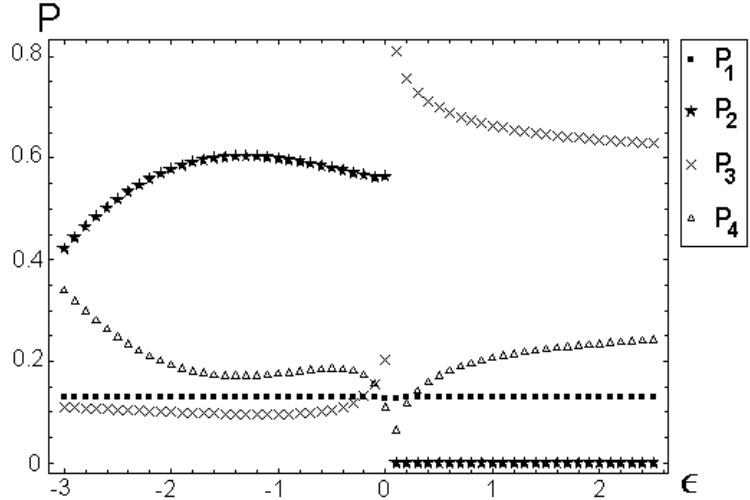}
\caption{Transition probabilities to all states in the model in Fig.\ref{LZfig4} as functions of the distance
$\epsilon$ between levels 2 and 1. Long before level crossings the probability of the state 1 is set to unity. The transition probabilities
are represented correspondingly by boxes for the state 1, stars for the state 2, crossed lines for the state 3 and triangles for the state 4.}
\label{LZfig5}
\end{figure}

As another example, consider a 4-state model shown in Fig.\ref{LZfig4}. 
Obviously, for $\epsilon >0$ the transition from the state 1 to the state 2 
is counterintuitive but for $\epsilon <0$ it is not. Fig.\ref{LZfig5} shows numerically calculated final probabilities to find the system in all 4 
states for the evolution from $t=-600$ to $t=600$ when initially only the state 1 is populated. 

One can see that the probability to remain in the state 1 does not depend on $\epsilon$, in agreement with the Brundobler-Elser conjecture. 
A tiny diviation from the Brundobler-Elser formula can be seen for two points with $\epsilon$ closest to zero. However, this should be explained as
due to the fact that $\epsilon=0$ is the critical point and it takes much more time for probabilities to saturate in its vicinity, but
in simulations the time interval had to be finite.  At negative
$\epsilon$ all other probabilities strongly depend on the distance between states 1 and 2. This can be explained partly
even in the independent crossing approximation as due to the interference among different semiclassical paths leading to the same final state. However,
for $\epsilon >0$ the independent crossing approximation does not predict any dependence of probabilities on $\epsilon$ if initially only the state
1 is populated. Nevertheless,
one can deduce from the Fig.\ref{LZfig5} that in addition to the state 1 only the transition probabilities to the state 2 become flat and have
indistinguishable from zero magnitudes in agreement with (\ref{eq3}). Transition probabilities to states 3 and 4 strongly depend on $\epsilon$ there. 
This indicates that simulations were performed for the range of parameters where the independent crossing approximation fails unless its predictions
become for some reasons exact. 
 
 In conclusion, the generalization of the Brundobler-Elser hypothesis is proposed that states that counterintuitive transitions in the multistate
  Landau-Zener model with the linear time-dependence of diabatic energies are asymptotically forbidden.
   It is confirmed by all numerical tests and by all known exact solutions. In addition, we demonstrated that this result can be 
   explained by continuation of the time-path into the complex plain, i.e. by the same approach as the one proposed by Landau to solve the two state model.
 As in any known exact solution of the multistate Landau-Zener model,  the formula (\ref{eq3}) coincides with predictions of the independent 
 crossing approximation. This
 fact points to the common origin of all such exact results.

\textit{Acknowledgments}. Author is grateful to V.L. Pokrovsky for
useful discussions. This work was supported by NSF under the grant
DMR0072115 and DMR 0103455 and by Telecommunication and Informatics Task Force at Texas A\&M University.



\begin{references}
\bibitem{Landau} L.D. Landau, Physik Z. Sowjetunion \textbf{2}, 46 (1932)

\bibitem{Zener} C. Zener, Proc. Roy. Soc. Lond. A \textbf{137}, 696 (1932)
\bibitem{Gefen} Y. Gefen and D. J. Thouless, Phys. Rev. Lett. 59, 1752-1755 (1987), K. Mullen, E. Ben–Jacob, Y. Gefen, and Z. Schuss, 
                Phys. Rev. Lett. 62, 2543-2546 (1989), D. Lubin, Y. Gefen, and I. Goldhirsch, Phys. Rev. B 41, 4441-4455 (1990) 



\bibitem{demkov} Yu. N. Demkov, V. I. Osherov, Zh. Exp. Teor. Fiz. 53 (1967) 1589 (Engl. transl. 1968 Sov. Phys.-JETP 26, 916)
\bibitem{zeeman2} C.E. Carroll, F. T. Hioe, J. Phys. A: Math. Gen. 19, 1151-1161 (1986)
\bibitem{brund} S. Brundobler, V. Elser, J. Phys. A: Math.Gen. 26 (1993) 1211-1227
\bibitem{bow} V.N. Ostrovsky, H. Nakamura, J.Phys A: Math. Gen. 30 6939-6950(1997)
\bibitem{dem33} Y.N. Demkov, V.N. Ostrovsky, J. Phys. B. 34 (12), (2001) 2419-2435
\bibitem{usuki} T. Usuki, Phys.Rev.B 56, 13360 (1997)
\bibitem{sinitsyn} N.A. Sinitsyn, Phys.Rev.B66, 205303 (2002)
\bibitem{math} Takashi Aoki, Takahiro Kawai and Yoshitsugu Takei  J. Phys. A: Math. Gen. 35 (2002) 2401-2430
\bibitem{shytov} A.V. Shytov, cond-mat/0312011
\bibitem{Yur}  V.A. Yurovsky, A. Ben-Reuven,  P.S. Julienne, Y.B. Band, J. Phys. B: At. Mol. Opt. Phys. 32 (1999) 1845

\end {references}
\end{document}